# Comments on "Quasiparticle self-consistent GW theory of III-V nitride semiconductors: Bands, Gap bowing, and effective masses" [Phys. Re. B 82, 115102 (2010)]

D. Bagayoko, L. Franklin, and G. L. Zhao
Department of Physics, Southern University and A&M College
Baton Rouge, Louisiana 70813, USA

## Abstract

An oversight of several previous results from local density approximation (LDA) calculations appear to have led to an incomplete, and hence misleading, characterization of the capability of density functional theory (DFT) to describe correctly the electronic properties of wurtzite GaN (w-GaN) and InN (w-InN) [Phys. Rev. B 82, 115102 (2010)]. These comments are aimed at presenting a different picture of the above capability for DFT calculations that solve self-consistently the system of equations of DFT. They also underscore, in light of the experimentally established Burstein-Moss effect, the need to specify the carrier density when citing a band gap for w-InN.

Svane et al.[1] reported results of their sophisticated quasiparticle self-consistent GW (QSGW) calculations for w-GaN, w-InN, and InGaN$_2$. These authors employed the linear muffin-tin orbital (LMTO) method in a full potential approach. All scalar-relativistic effects were taken into account in the QSGW self-consistency cycle. The authors also applied an empirical hybrid approach that combines 80% of the QSGW self-energy with 20% of LDA self-energy; this approach was reported to correct the slight overestimation of band gaps calculated within the QSGW approximation. These efforts resulted in electronic properties of w-GaN and w-InN that are much closer to corresponding measured values. In particular, the band gaps from their hybrid approach are practically the same as the corresponding experimental values, as shown in Table 1 of Reference 1. A similar statement holds for several values of electron effective masses where both the QSGW and hybrid results agree with experiment.

Svane et al.[1] also performed LDA calculations whose results were compared to those of the QSGW and hybrid calculations. Their LDA findings, as per the content of their Table 1, indicate that w-InN is a metal and that the fundamental gap[1] of w-GaN is 1.90 eV, a value quite different from the cited experimental ones of 3.50 and 3.51 eV. In light of these LDA results, Svane et al.[1] justifiably noted a failure of LDA to describe correctly the electronic properties of w-GaN and w-InN, with emphasis on the band gap. Citing some other GW calculations, the authors found what they consider to be a confirmation of a considerable improvement of GW calculated band gaps as compared to LDA results.

The aim of these comments is to note several previous LDA results for w-GaN and w-InN that are in agreement with experiment and hence contradict the above picture of failure of LDA, on the one hand, and of LDA as compared to GW, on the other. The oversight of these results, we believe, led to an incomplete and misleading characterization of the capability of LDA and DFT to describe correctly the electronic properties w-GaN and w-InN.

The content of the table below (Table 1) suffices to support our contention. The numbers in this table are as they appear in Reference 1, except for those in the two columns labeled LDA-BZW. The contents of these columns are from several previous LDA calculations[2-6] missed by

Reference 1. These LDA-BZW results were obtained from *ab-initio* calculations that solved self-consistently both the Kohn-Sham equation and the equation giving the ground state charge density in terms of the wave functions of the occupied states. This double self-consistency is obtained in calculations that utilize the Bagayoko, Zhao, and Williams (BZW).[2,6] Such calculations correctly solve the relevant *system of equations* defining LDA, as stated by Kohn and Sham[7] and reiterated by Kohn in his Nobel Lecture.[8]

The LDA-BZW band gap for w-GaN, 3.2 eV, is not any lower than the experimental value cited above than the QSGW gap of 3.81 eV is higher. While the LDA band gap of w-InN, as obtained by Svane et al.,[1] is negative, the LDA-BZW result of 0.88 is in excellent agreement with experiments[9-10] on w-InN samples with relatively high carrier densities. Indeed, with carrier densities of $1.2 \times 10^{19}$ cm$^{-3}$ and $5 \times 10^{19}$ cm$^{-3}$, Wu et al.[9] and Inishima et al.[10] reported w-InN band gaps of 0.883 and 0.889 eV, respectively. It should be noted that these values are in excellent agreement with the QSGW result of 0.99 eV. The experimental confirmation[11] of both our predicted lattice constant and band gap for cubic InN (c-InN),[4] using LDA potentials, is another indication of our contention that (a) LDA calculations that properly solve the required[7-8] *system of equations* provide a good description of the electronic properties of most semiconductors, in general, and of w-GaN, w-InN, and c-InN in particular and (b) that other approaches, including GW calculations, have yet to provide better results from first principle. Zhao et al.[2] and Bagayoko and Franklin[3] reported electron effective masses that are also in agreement with experiments, adding to the alternative picture of the capability of LDA as illustrated above.

In summary, data in Table 1 and the above discussions established the fact that correctly performed LDA or DFT calculations, as done in the Bagayoko, Zhao, and Williams method, provide very good descriptions of properties of w-GaN, w-InN, and c-InN. The many DFT or LDA corrections schemes that are growing in number have yet to lead to first principle results that are in better agreement with experiment than those from LDA calculations that self-consistently solve the relevant system of equations. The same statement holds for other non-LDA approaches. For these reasons, this other picture of the rather significant capability of LDA calculations to describe the above materials needed to be contrasted with the one that pervades Reference 1.

**Acknowledgments**

This work was funded in part by the Louisiana Optical Network Initiative (LONI, Award No. 2-10915), the Department of the Navy, Office of Naval Research (ONR, Award Nos. N00014-08-1-0785 and N00014-04-1-0587), and the National Science Foundation (Award Nos. 0754821, EPS-1003897, and NSF (2010-15)-RII-SUBR).

**Table 1.** Selected contents of Table 1 of Reference 1, with the addition of LDA-BZW results in Columns 5 and 9 for wurtzite GaN and InN, respectively. Results in columns labeled QSGW, Hyb., and LDA are from Reference 1. The energies are in electron volt (eV). Eg and $W(N_p)$ are the minimum band gap and the width of the group of upper valence bands, respectively.

|  | w-GaN | | | | w-InN | | | |
|---|---|---|---|---|---|---|---|---|
|  | QSGW[a] | Hyb.[a] | LDA[a] | LDA-BZW[b] | QSGW[a] | Hyb.[a] | LDA[a] | LDA-BZW[c] |
| Eg(min) | 3.81 | 3.42 | 1.90 | 3.2 | 0.99 | 0.74 | -0.21 | 0.88 |
| $W(N_p)$ | 7.6 | 7.53 | 7.14 | 7.1 | 6.07 | 6.02 | 5.91 | 5.61 |
| E(d) | -16.8 | -16.2 | -13.3 | -16 | -15.7 | -15.2 | -13.1 | -15.7 |
| $K_v$ | -3.07 | -3.02 | -2.76 | -2.8 | -2.38 | -2.33 | -2.20 | -2.14 |
| $K_c$ | 6.89 | 6.49 | 5.00 | 5.3 | 6.20 | 5.92 | 4.83 | 6.74 |
| $M_v$ | -1.1 | -1.07 | -1.01 | -1.1 | -0.96 | -0.95 | -0.91 | -0.87 |
| $M_c$ | 7.07 | 6.62 | 4.90 | 5.1 | 5.50 | 5.15 | 3.79 | 4.75 |

[a]Reference 1.   [b]Reference 2.   [c]References 3, 5, and 6.